\documentclass[dvips]{article}
\usepackage[dvips]{graphicx}
\usepackage{epsfig}
\usepackage{amsfonts}
\usepackage{amsmath}
\usepackage{mathrsfs}
\usepackage{amssymb}
\usepackage{makeidx}
\usepackage{cancel}

\newtheorem{theorem}{Theorem}[section]
\newtheorem{definition}{Definition}[section]

\numberwithin{equation}{section}

\author{Jeffrey M. Groah}
\begin{document}
{\setlength\arraycolsep{1pt} 
\title{$C^{1,1}$ Pseudohermitian, Torsion-free Manifolds}

\maketitle

\abstract{Riemannian Manifolds may be $C^{1,1}$ and the geometry of these manifolds is investigated in \cite{Groah1}. Here, a similar analysis is given for pseudohermitian, torsion-free manifolds whereby, instead of assuming that the metric is parallel, it is assumed that the metric is pseudohermitian, a condition adopted by Einstein and elaborated upon in \cite{Hlavaty}. At the level of regularity assumed here, Einstein's formulation of the pseudohermitian condition is not tensorial and so a reformulation of this condition is given here. It is shown that a $C^{1,1}$ manifold is pseudohermitian and torsion-free if and only if it is Riemannian.}

\section{Introduction}

The Einstein equations, as a system of second-order partial differential equations, are expected to produce metric solutions that are $C^{1,1}$ and hence spacetime coordinates that are $C^{2,1}$. This level of regularity is sufficient for the existence of locally inertial frames and in \cite{Re-Te} such coordinates are shown to exist in the case of shockwave interactions between shocks from different characteristic families. See also \cite{Israel} for the case of a single shockwave and \cite{Op-Sn, Sm-TeNon}. In \cite{Gr-Te}, existence of shock wave solutions to the Einstein equations is given in the case of spherically symmetric spacetimes. In the case of spherically symmetric solutions of the Einstein equations in the ansatz for the metric used in \cite{Gr-Te}, the metric components have only $C^{0,1}$ regularity, {\it i.e.,} are Lipschitz continuous. In such a spacetime, the coordinate functions are $C^{1,1}$ giving the manifold this level of regularity. This paper investigates the geometry of pseudohermitian spacetimes, to be defined below. 

We assume that we have a manifold $M$ that is $C^{1,1}$. The Jacobi Identity is altered and an extra term arises in spacetime torsion leading to additional terms in the connection and curvature. Failing to account for these terms introduces a nonzero torsion into spacetime, affects covariant derivatives and curvature computations, and introduces an extraneous acceleration into particle paths. Correcting for theses terms requires a re-evaluation of geometry from a foundational level.

\section{$C^{1,1}$ Geometry}

Define the commutator between coordinate vector fields by 
\begin{equation}
[\partial_{\lambda},\partial_{\mu}]=\frac{\partial^{2}}{\partial x^{\lambda}\partial x^{\mu}}-\frac{\partial^{2}}{\partial x^{\mu}\partial x^{\lambda}}.\label{CommDef}
\end{equation}
Normally, in differential geometry we require $[\partial_{\lambda},\partial_{\mu}]=0$ so that manifolds are endowed with the $C^{2}$ differential structure of $\mathbb{R}^{n}$. Clairaut's Theorem says that mixed second-order partials commute provided the first-order partials exist and are continuous in an open set and the second-order partials are continuous at the point in question. However, this paper assumes that coordinate functions are only $C^{1,1}$ and hence a re-evaluation of the principles of geometry is necessary. 

At this level of regularity second-order partials may not commute. In order to apply Frobenius' Integrability Condition \cite{Lee}, we must assume that $[\partial_{\lambda},\partial_{\mu}]$ is in involution and hence 
\begin{equation}
[\partial_{\lambda},\partial_{\mu}]=c_{\lambda\mu}{}^{\sigma}\partial_{\sigma},
\label{Involution}
\end{equation}
where we have used the Einstein summation convention whereby repeated up and down indices are summed over all permissible values. Note that $c_{\lambda\mu}{}^{\sigma}=-c_{\mu\lambda}{}^{\sigma}$ and that these coefficients depend on the nature of the irregularities in question. The assumption that $c_{\lambda\mu}{}^{\sigma}\equiv 0$ is equivalent to assuming that spacetime coordinates are $C^{2}$. Also, we use a 3-index notation whereby objects that have two lower indices and one upper index and are skew in the lower indices have the upper index on the right, those symmetric in the lower indices have the upper index on the left, and those that are neither have their upper index in the middle.

Note that $c_{\lambda\mu}{}^{\sigma}$ is not a tensor since $[fX,Y]=f[X,Y]-Y(f)X$. 

\begin{theorem}The quantities $c_{\lambda\mu}{}^{\nu}$ transform according to 
\begin{equation}
c_{ij}{}^{k}=c_{\alpha\beta}{}^{\nu}\frac{\partial x^{\alpha}}{\partial x^{i}}\frac{\partial x^{\beta}}{\partial x^{j}}\frac{\partial x^{k}}{\partial x^{\nu}}-\frac{\partial x^{\alpha}}{\partial x^{i}}\frac{\partial x^{\beta}}{\partial x^{j}}\left(\frac{\partial^{2} x^{k}}{\partial x^{\alpha}\partial x^{\beta}}-\frac{\partial^{2} x^{k}}{\partial x^{\beta}\partial x^{\alpha}}\right)\label{cTrans}
\end{equation}
and also
\begin{equation}
c_{ij}{}^{k}=c_{\lambda\mu}{}^{\nu}\frac{\partial x^{\lambda}}{\partial x^{i}}\frac{\partial x^{\mu}}{\partial x^{j}}\frac{\partial x^{k}}{\partial x^{\nu}}+\frac{\partial x^{k}}{\partial x^{\delta}}\left(\frac{\partial^{2} x^{\delta}}{\partial x^{i}\partial x^{j}}-\frac{\partial^{2} x^{\delta}}{\partial x^{j}\partial x^{i}}\right).\label{cTrans2}
\end{equation}
\end{theorem}

{\bf Proof.} Equation (\ref{cTrans2}) follows by direct substitution into (\ref{CommDef}) and applying (\ref{Involution}). Equation (\ref{cTrans}) follows from (\ref{ConnTrans}) or from (\ref{cTrans2}) and a trivial computation.\ $\Box{}$

\begin{definition}
The Jacobi tensor is defined by 
\begin{eqnarray}
J(X,Y,Z)&=&[X,[Y,Z]]+[Z,[X,Y]]+[Y,[Z,X]]\nonumber\\
&=&\sum_{\circlearrowleft}[X,[Y,Z]]
\label{Jacobi}
\end{eqnarray}
where $\sum_{\circlearrowleft}$ is the sum over cyclic permutations of the vector fields.
\end{definition}

That $J(X,Y,Z)$ is a tensor follows by a short computation. 

\begin{theorem}\label{JacIdent} The Jacobi tensor vanishes when $c_{\alpha\beta}{}^{\sigma}\equiv 0$ and 
\begin{equation}
J_{\alpha\beta\gamma}{}^{\sigma}=(c_{\alpha\beta}{}^{\sigma})_{,\gamma}+c_{\alpha\beta}{}^{\delta}c_{\gamma\delta}{}^{\sigma}+(c_{\gamma\alpha}{}^{\sigma})_{,\beta}+c_{\gamma\alpha}{}^{\delta}c_{\beta\delta}{}^{\sigma}+(c_{\beta\gamma}{}^{\sigma})_{,\alpha}+c_{\beta\gamma}{}^{\delta}c_{\alpha\delta}{}^{\sigma}.
\label{JIdent}
\end{equation}
\end{theorem}

{\bf Proof:} The proof follows by applying (\ref{Involution}) to the coordinate representation of (\ref{Jacobi}).\ $\Box{}$

Note that the Jocobi tensor may require interpretation in the sense of distributions. Also, the failure of the Jacobi tensor to vanish has important implications for the differential geometry of $C^{1,1}$ manifolds. The nature of this failure may include delta functions, though regardless the failure occurs on a set of measure zero. 

The components of the connection satisfy 
\begin{equation}
\nabla_{\partial_{\lambda}}\partial_{\mu}=\Gamma_{\mu\ \lambda}^{\phantom{\mu}\sigma}\partial_{\sigma}
\label{ConnDef}
\end{equation}
where special note must be taken concerning index location and spacing.

\begin{theorem}So that covariant derivatives yield tensors, the components of the connection must transform according to
\begin{eqnarray}
\Gamma_{i\ j}^{\ k}&=&\Gamma_{\mu\ \lambda}^{\phantom{\mu}\nu}\frac{\partial x^{\mu}}{\partial x^{i}}\frac{\partial x^{\lambda}}{\partial x^{j}}\frac{\partial x^{k}}{\partial x^{\nu}}-\frac{\partial x^{\lambda}}{\partial x^{j}}\frac{\partial x^{\mu}}{\partial x^{i}}\frac{\partial^{2} x^{k}}{\partial x^{\lambda}\partial x^{\mu}}\nonumber\\
&=&\Gamma_{\mu\ \lambda}^{\phantom{\mu}\nu}\frac{\partial x^{\mu}}{\partial x^{i}}\frac{\partial x^{\lambda}}{\partial x^{j}}\frac{\partial x^{k}}{\partial x^{\nu}}+\frac{\partial x^{k}}{\partial x^{\delta}}\frac{\partial^{2} x^{\delta}}{\partial x^{j}\partial x^{i}}.\label{ConnTrans}
\end{eqnarray}
\end{theorem}

{\bf Proof:} This is a short computation.\ $\Box{}$

Note that since second-order partials do not commute, care must be taken with respect to the order of the indices in (\ref{ConnTrans}). 

\begin{definition} Any object with two lower indices and one upper index that transforms according to the pattern in (\ref{ConnTrans}) is called a {\it connection}.
\end{definition}

\begin{definition}
The torsion tensor is defined by 
\begin{equation}
T(X,Y)=\nabla_{X}Y-\nabla_{Y}X-[X,Y].
\label{TorDef}
\end{equation}
\end{definition}
That (\ref{TorDef}) is a tensor follows by a short computation.

Coordinates for the torsion tensor are given by 
\begin{equation}
T_{\lambda\mu}{}^{\nu}=\Gamma_{\mu\ \lambda}^{\ \nu}-\Gamma_{\lambda\ \mu}^{\ \nu}-c_{\lambda\mu}{}^{\nu}
\label{TorComp}
\end{equation}
and are skew in their lower indices, $T_{\lambda\mu}{}^{\nu}=-T_{\mu\lambda}{}^{\nu}$.

\begin{theorem} For $C^{1,1}$ manifolds, torsion-free connections satisfy
\begin{equation}
\Gamma_{\mu\ \lambda}^{\ \alpha}=\Gamma_{\lambda\ \mu}^{\ \alpha}+c_{\lambda\mu}{}^{\alpha}.
\label{ChSym}
\end{equation}
\end{theorem}

{\bf Proof:} Torsion-free connections satisfy
\begin{equation}
T(X,Y)=\nabla_{X}Y-\nabla_{Y}X-[X,Y]=0,
\nonumber
\end{equation}
which will be affected by the integrability conditions (\ref{Involution}). This means that 
\begin{eqnarray}
0&=&\nabla_{\partial_{\lambda}}\partial_{\mu}-\nabla_{\partial_{\mu}}\partial_{\lambda}-[\partial_{\lambda},\partial_{\mu}]\nonumber\\
&=&\Gamma_{\mu\ \lambda}^{\ \alpha}\partial_{\alpha}-\Gamma_{\lambda\ \mu}^{\ \alpha}\partial_{\alpha}-c_{\lambda\mu}{}^{\alpha}\partial_{\alpha}
\nonumber
\end{eqnarray}
and hence (\ref{ChSym}) holds.\ $\Box{}$

{\it Nota bene:} Components of the connection are not necessarily symmetric in their lower indices, and 
\begin{equation}
\Gamma_{\lambda\ \mu}^{\ \alpha}-\Gamma_{\mu\ \lambda}^{\ \alpha}=-c_{\lambda\mu}{}^{\alpha}.\label{ChDiff}
\end{equation}

\section{The Pseudohermitian Condition}

The pseudohermitian condition was investigated by Einstein, and a development of the corresponding theory may be found in \cite{Hlavaty}. In Einstein's analysis, the metric was not assumed to be symmetric. Here, the metric is assumed to be symmetric. 

\begin{definition}
A metric $g_{\lambda\mu}$ is called {\rm parallel} if 
\begin{equation}
0=\partial_{\omega}g_{\lambda\mu}-\Gamma_{\lambda\ \omega}^{\phantom{\lambda}\sigma}g_{\sigma\mu}-\Gamma_{\mu\ \omega}^{\phantom{\lambda}\sigma}g_{\lambda\sigma}.\label{Parallel}
\end{equation}
\end{definition}

In \cite{Hlavaty}, a metric $g_{\lambda\mu}$ is called {\rm pseudohermitian} if 
\begin{equation}
0=\partial_{\omega}g_{\lambda\mu}-\Gamma_{\lambda\ \omega}^{\phantom{\lambda}\sigma}g_{\sigma\mu}-\Gamma_{\omega\ \mu}^{\phantom{\lambda}\sigma}g_{\lambda\sigma}.\label{Pseudo}
\end{equation}
Take special note of the indices of the connection components in the last term for these two definitions. Since the connection components are not symmetric in the lower indices, condition (\ref{Pseudo}) yields a geometry distinct from the Riemannian geometry determined by (\ref{Parallel}).

In the present context of $C^{1,1}$ manifolds, the pseudohermitian condition cannot be defined by (\ref{Pseudo}) since the expression on the right-hand-side is not tensorial at this level of regularity. In cognizance of (\ref{ChSym}), we must alter the definition in a way that is manifestly tensorial as follows:
\begin{definition}
A metric $g_{\lambda\mu}$ is called {\rm pseudohermitian} if 
\begin{eqnarray}
0&=&D_{\partial_{\omega}}(g)(\partial_{\lambda},\partial_{\mu})+g(\partial_{\lambda},T(\partial_{\omega},\partial_{\mu}))\nonumber\\
&=&\partial_{\omega}g_{\lambda\mu}-\Gamma_{\lambda\ \omega}^{\phantom{\lambda}\sigma}g_{\sigma\mu}-\Gamma_{\omega\ \mu}^{\phantom{\lambda}\sigma}g_{\lambda\sigma}-c_{\mu\omega}{}^{\sigma}g_{\lambda\sigma}.\label{Pseudo2}
\end{eqnarray}
\end{definition}

It immediately follows that a manifold is pseudohermitian and torsion-free if and only if it is Riemannian, {\it i.e.,} the metric is parallel and the torsion is identically zero. From this, all of the analysis in \cite{Groah1} holds for pseudohermitian, torsion-free manifolds.

\section{Conclusion} At the level of $C^{1,1}$ regularity, the definition of pseudohermitian manifolds found in \cite{Hlavaty} fails to be tensorial and hence requires modification. A manifestly tensorial definition is given here from which it immediately follows that pseudohermitian, torsion-free manifolds are Riemannian. It follows that all of the analysis found in \cite{Groah1} applies to the present case as well.

\end{document}